\documentclass[reprint,amsmath,amssymb,aps]{revtex4-1}
\usepackage[utf8]{inputenc}
\usepackage{CJK}

\usepackage{graphicx}
\usepackage{dcolumn} 
\usepackage{bm} 
\usepackage{physics}

\begin{document}
\begin{CJK*}{GB}{}

\title{Assisted percolation of slow-spreading mutants in heterogeneous environments}

\author{Thomas Tunstall}
\altaffiliation{Living Systems Institute, Faculty of Health and Life Sciences, University of Exeter}
\altaffiliation{Physics and Astronomy, Faculty of Environment, Science and Economy, University of Exeter}

\author{Tim Rogers}
\altaffiliation{	
Center for Networks and Collective Behaviour, Department of Mathematical Sciences, University of Bath}

\author{Wolfram M\"obius}
\altaffiliation{Living Systems Institute, Faculty of Health and Life Sciences, University of Exeter}
\altaffiliation{Physics and Astronomy, Faculty of Environment, Science and Economy, University of Exeter}

\date{\today}

\begin{abstract}
Environmental heterogeneity can drive genetic heterogeneity in expanding populations; mutant strains may emerge that trade overall growth rate for an improved ability to survive in patches that are hostile to the wild type. This evolutionary dynamic is of practical importance when seeking to prevent the emergence of damaging traits. We show that a sub-critical slow-spreading mutant can attain dominance even when the density of patches is below their percolation threshold and predict this transition using geometrical arguments. This work demonstrates a phenomenon of ``assisted percolation'', where one sub-critical process assists another to achieve super-criticality.
\end{abstract}

\maketitle
\end{CJK*}

Pesticides are used to control crop pests, antimicrobials to eliminate microbes, and cancer drugs to contain tumours. The emergence of mutants that are resistant to these agents is a major concern in all these scenarios \cite{philbert2014review,ramakrishnan2019local,mansoori2017different}. The double-edged sword of use and control on the one hand and loss of efficacy through the emergence of resistance is widely acknowledged, but we know relatively little about the role spatial structure plays in the dynamics of resistance emergence. Research on the effects of compartmentalisation, for example in the human body \cite{moreno2015imperfect}, gradients mimicking such compartmentalisation \cite{kepler1998drug}, and mosaic application, for example on fields of crops \cite{rimbaud2018}, have all emphasised the role of a reservoir with low concentrations of the control agent. Therein, mutants can originate that are then selected for in regions with a higher concentration. How these mutants then spread in a complex environment is not understood.

\begin{figure}[b!]
    \centering    \includegraphics{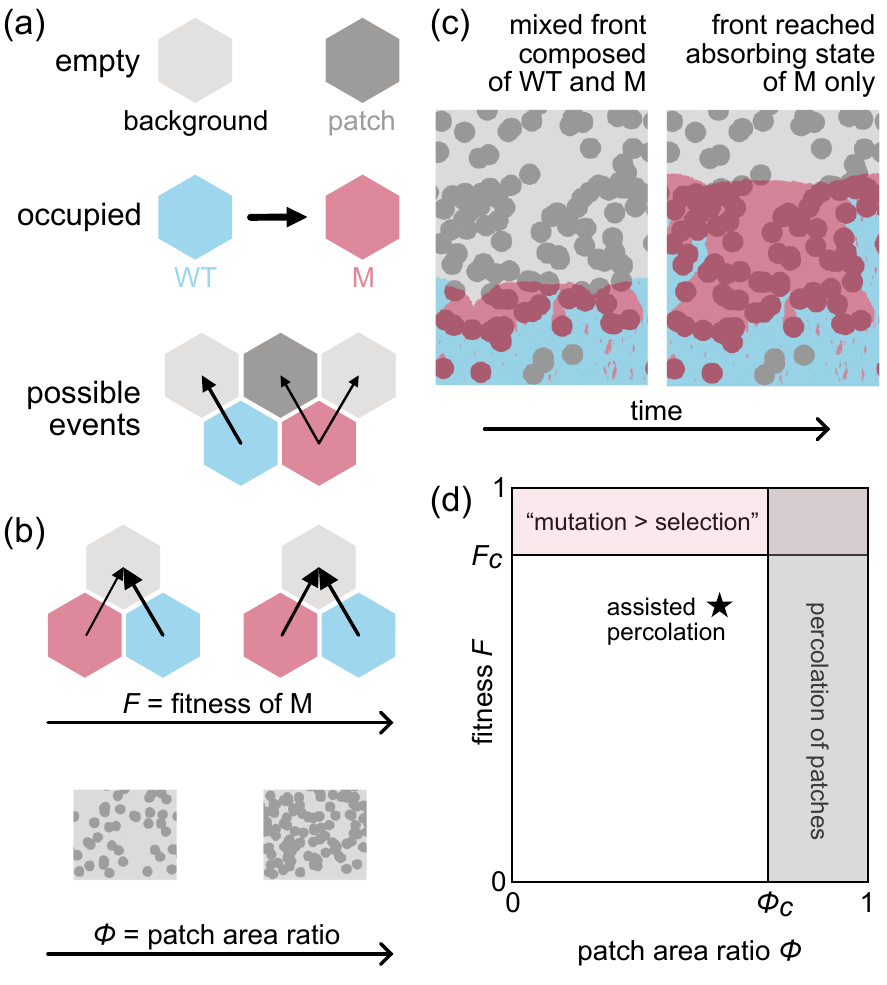}
    \caption{\textbf{(a)} Rules underlying the modified Eden model. The possible reproduction events of the next simulation step are represented by arrows. See also Fig.\ S1a. \textbf{(b)} Visualisation of the effects of increasing fitness $F$ of M individuals and patch area ratio $\phi$. See also Fig.\ S1b. \textbf{(c)} M dominates in an expanding front in the presence of treated patches (dark circles) in which only M (dark/red) survive. See Video 1 for a depiction of the full simulation. \textbf{(d)} Sketch of phase diagram, for sufficiently high fitness $F>F_c$ and patch area $\phi>phi_c$ dominance of M is expected. Parameters for simulation in panel (c) lie outside of these regions, indicating a richer structure to the phase diagram in which assisted percolation takes place.
    }
    \label{Fig1}
\end{figure}

We here address this question from a theoretical physics perspective, for a two-dimensional environment with isolated patches that can be thought of as being protected by the control agent. We generalize the Type C variant \cite{Jullien1985} of the Eden model \cite{Eden1961}, a lattice-based model for growth which is suitable for spread in heterogeneous environments, intrinsically incorporates stochasticity, and whose computational efficiency matches the requirement to investigate large systems and many replicates (Figs.\ \ref{Fig1}a-b and S1, and expanding on previous work \cite{Korolev2010,Hallatschek2007,Moebius2015,gralka2019environmental,Moebius2018}, see Supplementary Material).

Incorporating environmental heterogeneity, mutation, and selection in a two-species Eden model, we consider a hexagonal lattice of sites. Unoccupied sites can be either `background' or `patch site' varieties, whilst occupied sites are either either `Wild-Type' (WT) or `Mutant' (M) pathogens (Fig.\ \ref{Fig1}a). During each time step an occupied site on the population frontier is selected to reproduce, at which point offspring of the same type as the parent are placed in a random unoccupied neighbouring site; only M are able to reproduce into a patch site, to represent their resistance to the control measure. During reproduction, there is a finite probability $\mu$ of mutation from WT to M (Figs.\ \ref{Fig1}a and S1a). Selection of WT or M sites to reproduce occurs with probability proportional to their `fitness' which is taken to be 1 for WT and $F<1$ for M, modelling the cost of resistance \cite{fisher1999genetical} (Fig.\ \ref{Fig1}b).

In finite-width spreading fronts, M will eventually come to dominate as this is the only absorbing state. However, the timescale can vary dramatically: Even in the absence of patch sites, one can distinguish a super-critical phase of fast fixation and a sub-critical phase of exponentially slow fixation \cite{Kuhr2011}. Below criticality, small clusters of M appear, but typically die out before coalescing with others. Increasing M fitness or mutation rate causes these clusters to grow in size or frequency, respectively, to the point where multiple coalescence events can occur and the M population becomes supercritical (Fig.\ S1b). For a similar model with a flat expanding front (Fig.\ S2a), the dynamics fall into the directed percolation universality class \cite{odor2004}. The roughness inherent to the Eden Model we use as the basis for our modified model greatly perturbs us from the DP universality class. \textit{Kuhr et al} \cite{Kuhr2011} performed phenomenological analysis on the Eden Model to determine the phase boundary at $\mu \approx p^{*} \left(1-F\right)^{1.4}$ with $p^*\approx0.407$ \cite{Kuhr2011}. For a given mutation rate, we can define the critical fitness $F_c\left(\mu\right)$. As a general result, M dominates quickly if $F>F_c(\mu)$.

We incorporate macroscopic environmental heterogeneity into this model by arranging treated patch sites into circular patches of fixed radii, either randomly placed (Fig.\ \ref{Fig1}b) or in a lattice arrangement. This type of heterogeneity differs from that of previous work by acting asymmetrically on different genotypes \cite{gralka2019environmental}: They act as hard boundaries for WT \cite{Moebius2021}, but are transparent to M. For random placements, continuum percolation theory gives a critical threshold of $\phi_c^\circ \approx 0.68$ \cite{xia1988} for the ratio $\phi$ of area covered by patches to total area; for $\phi>\phi_c^\circ$ the patches themselves form a percolating cluster allowing only M lineages to survive. Therefore, for a given mutation rate, M dominates either if $F>F_c$ or $\phi>\phi_c^\circ$. The example shown in Fig.\ \ref{Fig1}c demonstrates, however, that these are merely sufficient conditions. Rapid domination of M can occur with both fitness and patch area ratio being significantly lower than the critical thresholds. We aim to examine the full structure of the phase diagram sketched in Fig.\ \ref{Fig1}d.

Close examination of simulations such as that presented in Fig.\ \ref{Fig1}c and Video S1 reveals the mechanisms driving M dominating. When small M clusters intersect with a treated patch they spread through it and emerge from the other side ahead of the faster spreading WT population that is forced to take a longer route around the patch acting as an obstacle. If the `escape region' beyond a treated patch is large enough, it will intersect with another patch and M population growth will continue. This is an effect of ``assisted percolation" as we will explore later. To determine the boundary of the fast fixation phase, it is therefore necessary to compute (i) the expected size of escape regions, and (ii) the effective between-patch percolation process. 

\begin{figure}[b!]
    \centering
    \includegraphics{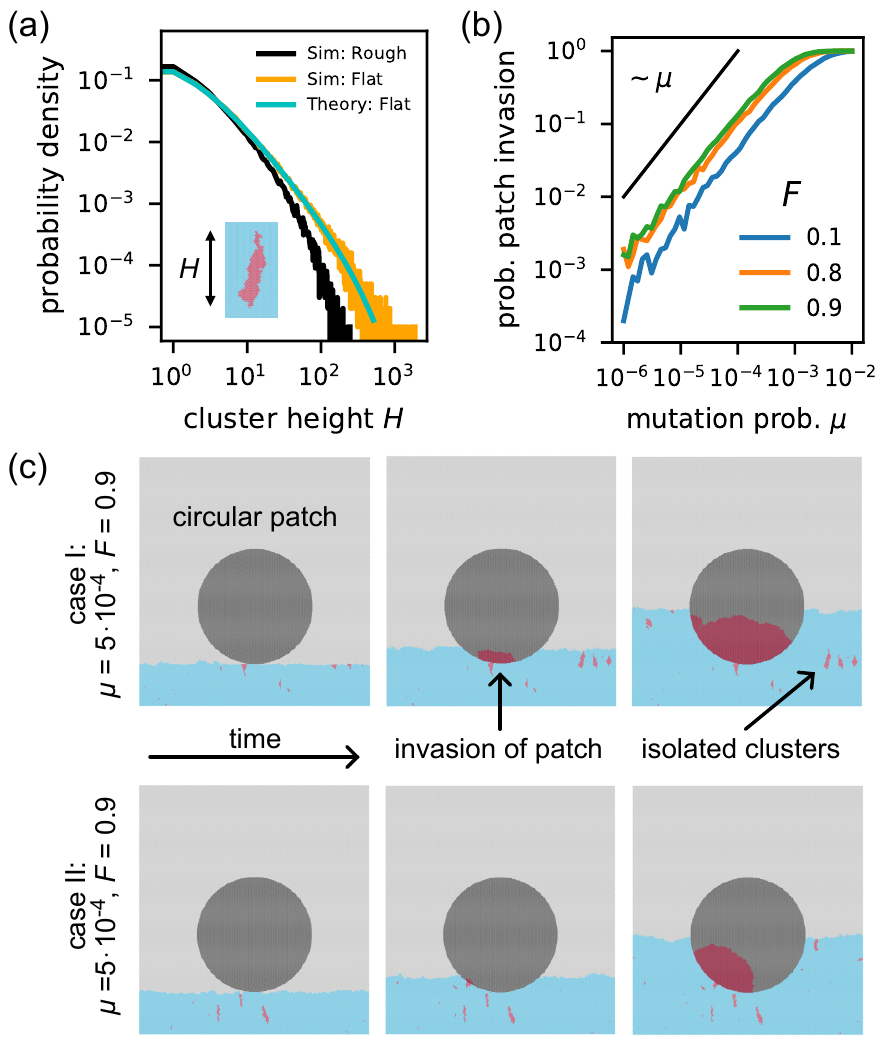}
    \caption{\textbf{(a)} Cluster height distributions for the Eden model (rough front) and a flat-front model, both obtained from simulations, together with the analytical result for flat fronts for $F=0.9$. Inset: Typical clusters in our model. \textbf{(b)} Probability of an isolated patch being invaded as a function of mutation rate for different fitness values $F$. Black line represents a linear relationship. \textbf{(c)} Two examples for how clusters invade a patch with $\mu= 5\times 10^{-4},\,F=0.9$: (case I) Invasion from the bottom and (case II) invasion from the bottom left. In both cases M inside the patch lags the WT front outside.}
    \label{Fig2}
\end{figure}

At first glance, the statistics of lone M clusters are important to this problem. These statistics are remarkably complex; to our knowledge, only the scaling behaviours have been determined for a square lattice in the literature \cite{Kuhr2011}. However, as demonstrated in Fig.\ \ref{Fig2}a, the vast majority of isolated M clusters in the regime considered here are much smaller than the typical size of the patches (area $\sim 10^3$). We have undertaken further analytical work in determining the dimensions of the lone M clusters for an equivalent model with a flat front (expanding on previous work \cite{KPZ,Kuhr2011,domany1984,essam1989,DyckPaths,WolframAlpha,oeisA080934,oeisA080936,gralka2018,lavrentovich2013}, see Supplementary Material).

How often these clusters lead to invasion of a patch depends on mutation rate and fitness. Fig.\ \ref{Fig2}b illustrates that the probability to invade a patch increases linearly and then saturates with increasing mutation rate as expected; similarly, a higher fitness results in higher probability of invasion. To abstract from both the sizes of isolated clusters and their abundance, we focus on the consequences of individual patches being invaded. In this way, we capture the long-term behaviour of the `thermodynamic limit' of a large system of patches with rare mutations.

\begin{figure}[b!]
    \centering
    \includegraphics{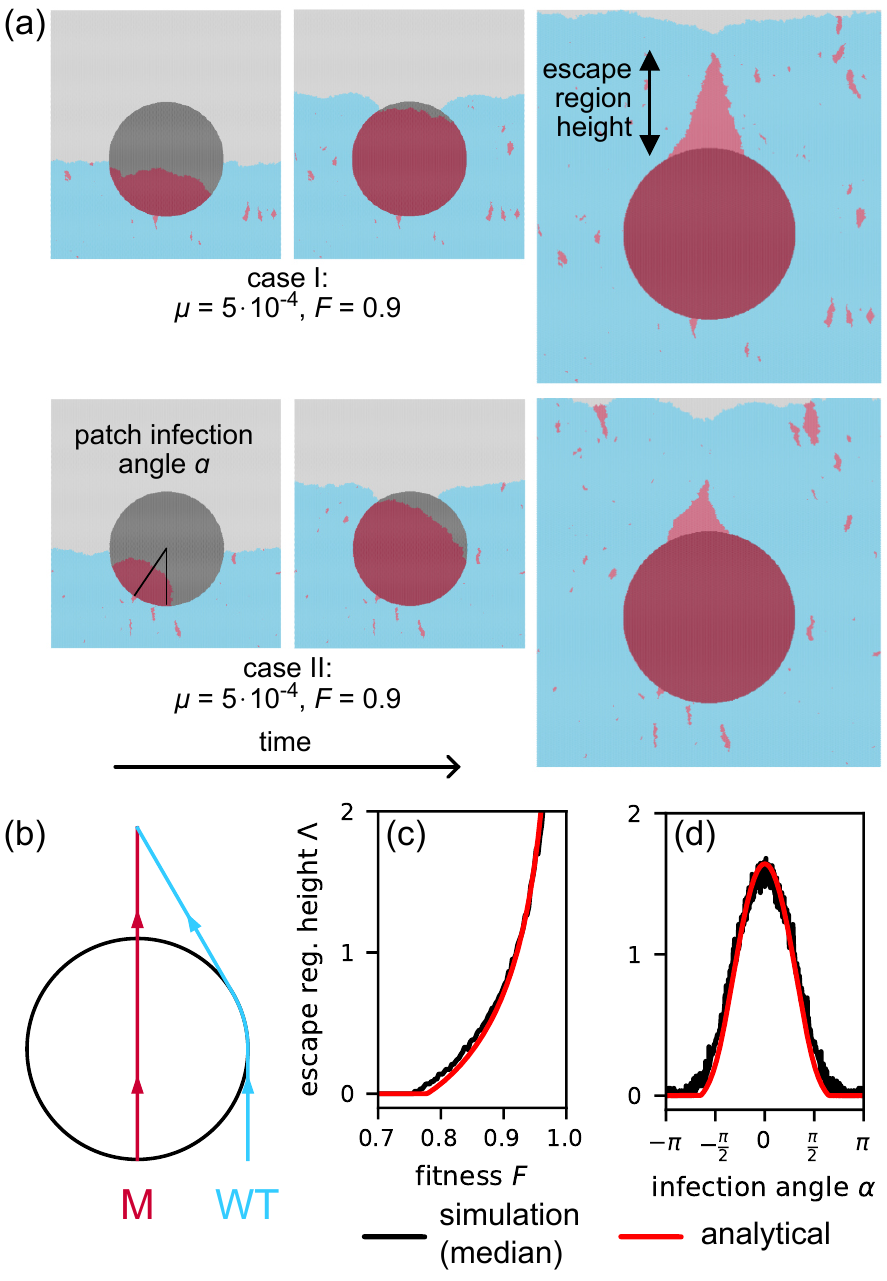}
    \caption{\textbf{(a)} Continuing the evolution of the patch invasions in Fig.\ \ref{Fig2}c with emphasis on the escape region. See Videos S2 and S3 for a depiction of the full simulation. \textbf{(b)} Sketch of the deterministic escape region height, found by equating the time taken for M to pass through the patch with the time taken for WT to pass around the patch to the same point. \textbf{(c)} Median of the normalised escape region height as a function of fitness $F$ for patch invasion angle $\alpha=0$. Black data points: Simulation results, red line: Geometrical prediction. \textbf{(d)} Like panel (c) but as a function of patch invasion angle $\alpha$ for fitness $F=0.95$.}
    \label{Fig3}
\end{figure}

While clusters are small relative to the patch's radius when outside a patch, they can spread unimpaired through the patch, leading to large domains within patches, as seen in Fig.\ \ref{Fig2}c. This domain may eventually become trapped within or escape the patch. Examples of the latter are shown in Fig.\ \ref{Fig3}a, where the invading M domain spreads upward through the patch, and is able to escape before it can be headed off by WT. We expect the existence of the escape region and, if applicable, its height to depend on the patch invasion angle $\alpha$ (which acts as the starting point for a race between M and WT strains), M's fitness $F$ (the relative speed of M), as well as stochastic effects. In fact, simulations with invasions seeded at different locations of the patch's boundary show that the median of escape region height increases with fitness $F$ and decreases with invasion angle, i.e., it is largest if the invasion occurs at the bottom of the patch (Fig.\ \ref{Fig3}c,d).

To understand this dependence quantitatively we turn to geometric arguments. Previous work has characterised front shape of a population encountering an obstacle in the absence of mutations and if front speed is the same everywhere outside the obstacle \cite{Moebius2015}. There, the front shape was determined as the set of all points that can be reached within a given time. Here we aim to find the point along the symmetry axis which is reached at the same time by WT expanding around the patch (with relative speed $1$) and M expanding through the patch (with relative speed $F$), Figs.\ \ref{Fig3}d and S3a,b. Measured in units of patch radius, we find that the typical maximum extent of escape regions $\Lambda\left(F,\alpha\right)$ solves the following equation (for $\abs{\alpha}<\frac{\pi}{2}$):
\begin{equation}
    \begin{split}
    &\frac{1}{F} \sqrt{1 + 2\left(1+\Lambda\right)\cos{\abs{\alpha}} +\left(1+\Lambda\right)^2} = \\ &\cos{\abs{\alpha}} + \arcsin{\frac{1}{1+\Lambda}} + \sqrt{\left(1+\Lambda\right)^2-1}
    \end{split}
    \label{Eq:EscapeRegionHeightBelow}
\end{equation}
Full details of the derivation are given in the Supplementary Material. If a real, positive solution does not exist, this means that the M were cut off immediately and did not escape, thus $\Lambda=0$. The numerical solution of Eq.\ \ref{Eq:EscapeRegionHeightBelow} describes the simulation data well when varying fitness $F$ or patch invasion angle $\alpha$ (Fig.\ \ref{Fig3}c,d). In the following, we will limit the discussion to patches invaded at the bottom and consequently the tallest escape regions, given that we expect the majority of patch invasions to occur around $\alpha\approx 0$.

\begin{figure}[b!]
    \centering\includegraphics{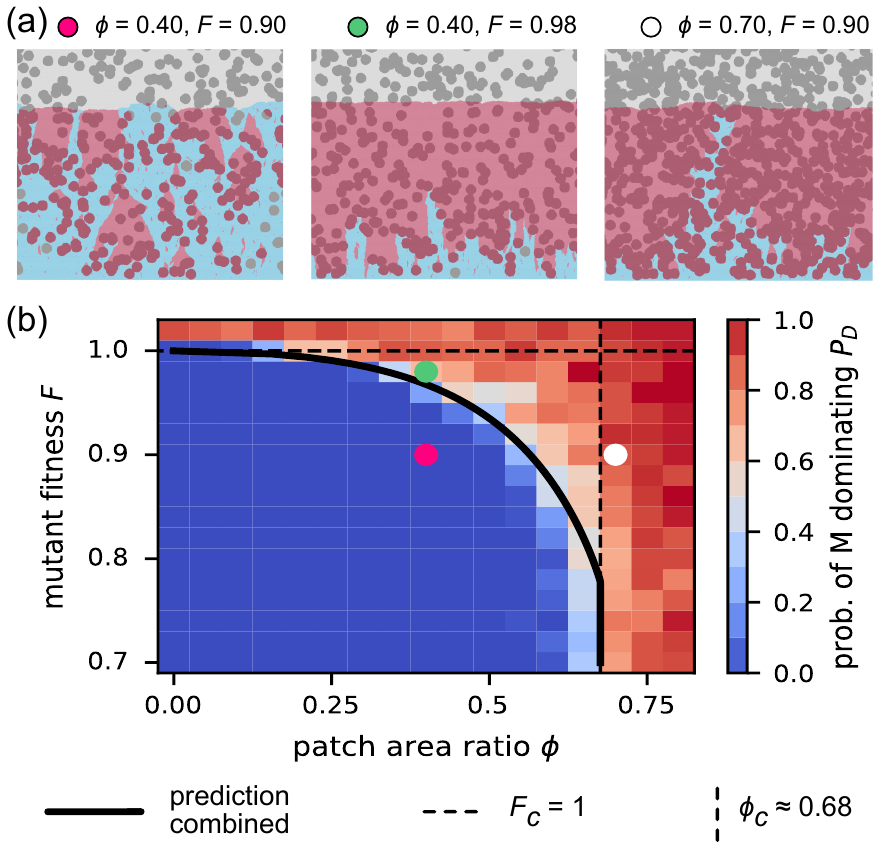}
    \caption{\textbf{(a)} Three snapshots of simulations for different values of fitness $F$ and patch area ratio $\phi$. See Videos S4-S6 for depictions of the full simulations. \textbf{(b)} Grid heat map of the probability of M dominating the front determined by simulations. The black line indicates the prediction of the boundary by Eq.\ \ref{Eq:Percolation}.}
    \label{Fig4}
\end{figure}

Having developed an understanding of the escape region from a single patch, we can examine the macrostructure emergent in a system of many randomly distributed patches. Three expansions for mutation rate $\mu=10^{-3}$ and with varying fitness $F$ and patch area ratio $\phi$ are displayed in Fig.\ \ref{Fig4}a. For low $F$ and $\phi$, patches and/or escape regions rarely overlap, while when either of these values crosses a threshold overlaps appear to lead to a growth of the fraction of M and ultimately domination of the front. To predict the patch area ratio at which this transition takes place, we estimate the percolation threshold for patches including escape regions (Figure S4): We treat the escape region (of rescaled height $\Lambda$) as a deformation to the patch shape, elongating in the direction of motion of the population front. A simple but effective heuristic is to consider M populations entering at the base of each patch, and treat the deformation as approximately elliptical. The percolation threshold for this system of ellipses can be found by rescaling the vertical direction to deform ellipses into disks. Conversely, the percolation threshold for a system of randomly distributed disks can be used to approximate the percolation threshold for the system of patches with escape regions, which is a function of fitness $F$ and patch area ratio $\phi$ and which we denote by $\phi^*_c(F)$. We reuse $\phi_c^\circ$ for the percolation threshold of disks and obtain (see Supplementary Material for details):
\begin{equation}
    \phi_c^*\left(F\right) = 1 - \left(1-\phi_c^\circ\right)^{\frac{2}{2 + \Lambda(F)}}\, ,
    \label{Eq:Percolation}
\end{equation}
where $\Lambda(F,\alpha=0)$ is the rescaled escape region height determined by Eq.\ \ref{Eq:EscapeRegionHeightBelow}. As expected, for vanishing escape region height, $\Lambda=0$, we obtain $\phi_c^*=\phi_c^\circ$, i.e., M will only dominate if patches themselves percolate. In this argument we demonstrate that super-criticality can be achieved via the dynamics of a sub-critical M population being perturbed by the presence of a sub-critical area ratio of patches, hence the term ``assisted percolation''.

To test how well Eqs. \ref{Eq:EscapeRegionHeightBelow} and \ref{Eq:Percolation} capture the transition from subcritical to supercritical regime, we simulated the system $50$ times for a wide range of fitness values $F$ and patch area ratios $\phi$ with patch radius $R=50$ and computed the probability $P_D$ with which M fixes at the front conditional on invasion of one patch (Fig.\ \ref{Fig4}, see Supplementary Material for details). To ensure that we only study the fate of a single mutation, the mutation rate is set to $0$ after a mutation occurs. To ensure that we have studied the case where M has invaded a patch, as the isolated cluster grows we keep track of the number of M on the population frontier: If this value ever exceeds the diameter of a patch, we can be confident that a patch has been invaded. If a cluster collapses before this threshold is met, the simulation is re-run for the same distribution of patches.  The transition region is characterised by $P_D$ being distinct from $0$ and $1$ and is indicated by lighter colors. $\phi_c^*\left(F\right)$, indicated as a black line, indeed captures this transition region very well. This means that not only the approximations made, but also the description of macrostructures interacting with each other capture the dynamics of the system very well.

\begin{figure}[b!]
    \centering \includegraphics{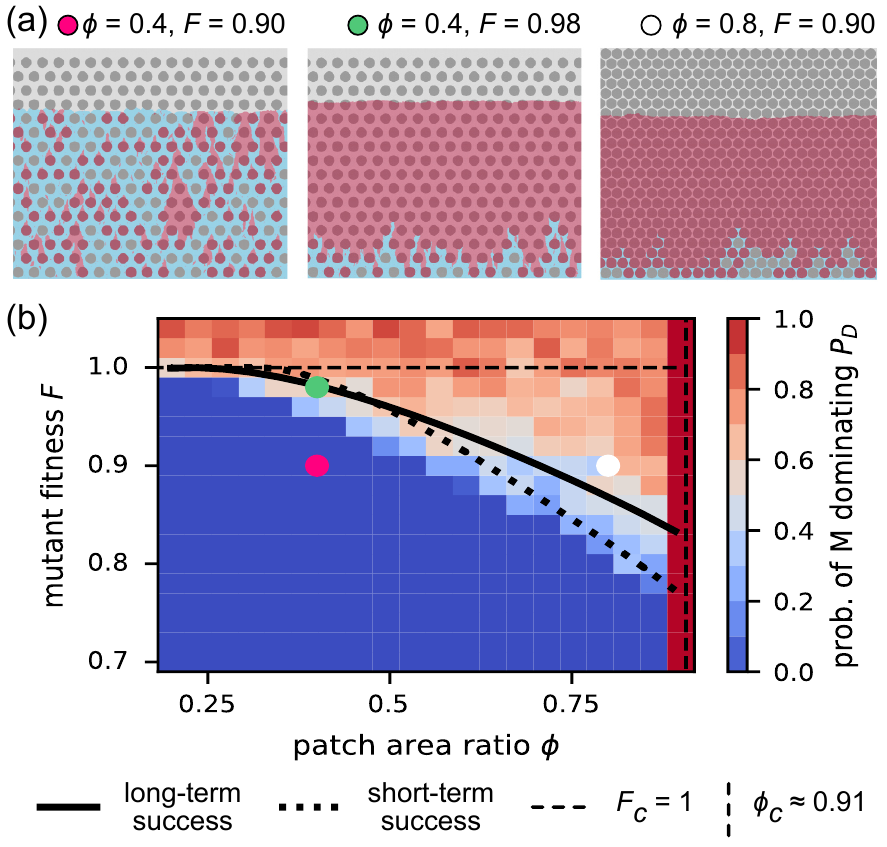}
    \caption{\textbf{(a)} Three snapshots of simulations for different combinations of $\phi$ and $F$. See Videos S7-S9 for depictions of the full simulations.  \textbf{(b)} Grid heat map of the probability of M dominating the front, determined by simulation. Dotted black line indicates a numerical prediction generated by consideration for short-time success, solid black line indicates analytical prediction generated by consideration for long-term success. Dashed lines indicate the mutation-fitness and patch coalescence phase transitions.}
    \label{Fig5}
\end{figure}

Motivated by wanting to further explore the applicability of these geometric arguments, and to develop a symmetrical patch distribution which can be designed to inhibit M domination, we considered patches organised on a hexagonal lattice. For a given patch radius $R$, the patch area ratio $\phi$ is a function of the lattice constant (the separation between the centres of adjacent patches). A sufficient condition for M dominating is $\phi$ to be larger than $\phi_{contact}\approx \frac{\pi}{2\sqrt{3}}$, at which point patches are in contact and thus not leave a path for WT to propagate, with the approximation capturing lattice artifacts. Fig.\ \ref{Fig5}a demonstrates that the phase transition profile permits rapid M domination below each of these thresholds, and we again computed the probability $P_D$ of M to dominate the front (Fig.\ \ref{Fig5}b) following invasion of one isolated patch (motivated by \cite{xia1988,torquato2002,chang2010simple}: See Supplementary Material for details).

To further characterise the region of fitness $F$ and patch area ratio $\phi$ within which M dominates the front quickly, we consider two different rationales, one valid on short, the other on longer time scales. For short times, we ask whether an escape region can lead to invasion of an adjacent downstream patch. For long times, we expect M to dominate if it can propagate faster vertically through the lattice of patches than WT. This transition can be determined by comparing the path length of WT snaking around patches while M passing straight through, similar to the computation of escape region height above, which yields the analytical result:
\begin{equation}
    F_c(\phi) = \frac{\sqrt{3}}{2} \frac{2+\eta\left(\phi\right)}{\sqrt{\eta\left(\phi\right)^2 +4\eta\left(\phi\right)} + 2\left(\frac{\pi}{3} - \arccos\frac{2}{2+\eta\left(\phi\right)}\right)}
    \label{Eq:LongTermSuccess}
\end{equation}
with $\eta\left(\phi\right)=\sqrt{\frac{2\pi}{\sqrt{3}\phi}}-2$. Both of these approaches are demonstrated in Fig.\ \ref{Fig5}b (and fully described by Fig.\ S5 as well as and Supplementary Material). As Fig.\ \ref{Fig5}b illustrates, the transition computed numerically for the short-term argument and the transition based on the analytical long-term argument yielding Eq.\ \ref{Eq:LongTermSuccess} adequately match the simulation results.

Comparing Fig.\ \ref{Fig4}b to Fig.\ \ref{Fig5}b demonstrates that the choice of patch distribution strongly affects the phase transition. Tackling the complex optimisation problem of preventing M domination for given patch area ratio would be a natural next step in the translation of this work to an applied setting. One would also need to incorporate a finite mutation rate, which we anticipate will perturb these results.

We mapped the question of how mutants spread in a complex environment of control agents to a modified Eden model with mutations in a heterogeneous environment. In the analysis, we incorporated results from disparate analyses of the Eden model (mutation-selection balance in the absence of patches \cite{Kuhr2011} and the perturbation of a single-strain front in the presence of obstacles \cite{Moebius2015}). Our observation that two sub-critical processes can combine to achieve super-criticality may have wider relevance: For example, one may consider the interaction between vaccine deployment and the emergence of vaccine-escape variants in epidemiology. The presence of a similar assisted percolation dynamic in social contact networks could have wide ramifications for the deployment of disease intervention strategies.

In this paper, we have mapped the question of how mutants spread in a complex environment of control agents to a modified Eden model with mutations in a heterogeneous environment. In doing so, we have demonstrated the existence of the novel dynamic of assisted percolation in a generalised version of a popular surface growth model. Although the model we have presented here is the first example of assisted percolation that we are aware of, we speculate that the phenomenon might be relevant to a range of other systems. In particular we expect that further examples may be found in the field of complex networks where, for example, a weak signal might achieve long-range transmission through a sub-critical set of amplifying nodes; potentially important applications to epidemiology and social dynamics are not hard to imagine.

\textbf{Acknowledgements:} Thomas Tunstall acknowledges support by EPSRC DTP and Syngenta Crop Protection. Wolfram M\" obius acknowledges support by BBSRC via BBSRC-NSF/BIO grant BB/V011464/1. Most of the simulations of this paper were performed on University of Exeter's high performance computer ISCA.

\bibliography{bibliography}

\end{document}


\title{Supplementary Material to ``Assisted percolation of slow spreading mutants in heterogeneous environments''}

\author[1,2]{Thomas Tunstall}
\author[3]{Tim Rogers}
\author[1,2]{Wolfram M\"obius}

\affil[1]{Living Systems Institute, Faculty of Health and Life Sciences, University of Exeter}
\affil[2]{Physics and Astronomy, Faculty of Environment, Science and Economy, University of Exeter}
\affil[3]{Center for Networks and Collective Behaviour, Department of Mathematical Sciences, University of Bath}

\maketitle

\tableofcontents

\section{Generalised Eden model}

\begin{figure}[h!]
    \centering\includegraphics{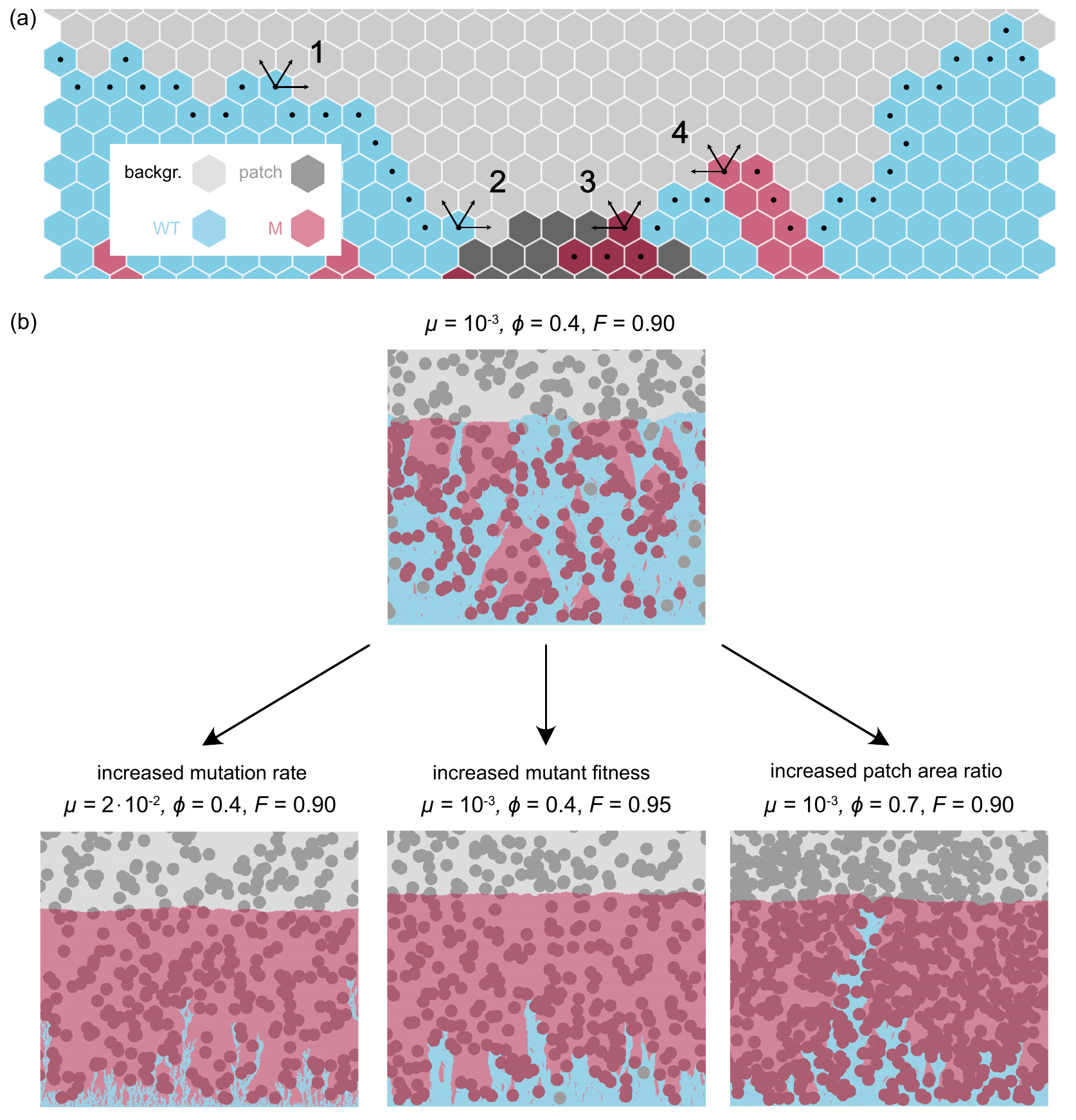}
    \caption{\textbf{(a)} (Inset) Rules underlying the modified Eden model. Each site is either a `patch site' or part of a `background' site. Occupied sites can be in of type `wild type' (WT) or `mutant' (M). Upon reproduction into an unoccupied neighboring site mutation from WT to M is possible (but not the reverse). Patch sites can only be turned into M sites. (Main panel) Section of the hexagonal lattice including the front of occupied sites. Sites that can convert a neighboring unoccupied site are marked with black dots. For sites marked 1-4, arrows indicate which neighboring unoccupied sites can be turned into occupied sites. \textbf{(b)} Snapshots of simulations for different sets of parameters for mutation rate $\mu$, patch area ratio $\phi$, and fitness $F$ . (Top) $\left(\mu=10^{-3},\phi=0.4,F=0.9\right)$: M and WT coexist, clusters of M typically disappear before merging into larger and larger clusters.  (Bottom, left) $\left(\mu=2\times10^{-2},\phi=0.4,F=0.9\right)$: Increasing the mutation rate results in more frequent clusters coalescing, the absorbing state of all M at the front is reached quickly. (Bottom, center) $\left(\mu=10^{-3},\phi=0.4,F=0.95\right)$: Increasing the mutant fitness results in the clusters getting larger, nearby clusters coalesce, the absorbing state of all M at the front is reached quickly. (Bottom, right) $\left(\mu=10^{-3},\phi=0.7,F=0.90\right)$: Increasing the patch area ratio provides a `geographical benefit' to M, the absorbing state of all M at the front is reached quickly.
    }
    \label{Fig1_Supplement}
\end{figure}

To investigate the dynamics of mutants spreading in a heterogeneous environment, we employ a generalisation of the Eden Model \cite{Eden1961}, type C \cite{Jullien1985}. The classical Eden model is a lattice model for surface growth \cite{Eden1961,Jullien1985}. Sites can be either occupied or empty. During the growth process in the Eden model of type C, a site with at least one empty neighboring site is randomly chosen to occupy or propagate one neighboring site. This mimics replication on a population front and suggests a way to implement mutations during this propagation process.

The original Eden model has been generalised to include occupied sites of different types, interpreted as genotypes in an expanding population \cite{Korolev2010,Hallatschek2007}. Modification of the lattice allows one to investigate the effects of environmental heterogeneity, such as inaccessible sites \cite{Moebius2015}, `disorder sites' of intermediate growth rate \cite{gralka2019environmental}, or surface curvature \cite{Moebius2018}.

In this work, we use a hexagonal lattice on which two different genotypes, denoted as wild type (WT, blue) and mutant (M, red), can propagate (Fig.\ \ref{Fig1_Supplement}a). M can progress onto all unoccupied sites on the lattice - the lattice is homogeneous from their perspective. WT, however, is barred from occupying hostile `patch sites' denoted in dark grey, and the environment is heterogeneous from WT's perspective. These patch sites are arranged into circles of radius $R$ (in units of lattice spacing), with area ratio $\phi$. During expansion, a site with neighbors that can be occupied (denoted in Fig.\ \ref{Fig1_Supplement}a by black dotted sites) is randomly chosen. In this process, WT sites are chosen with relative weight $1$ while M sites are chosen with weight (`fitness') $F<1$. One of the neighboring sites (black arrows in Fig.\ \ref{Fig1_Supplement}a) is chosen to be converted into the same type as the parent site. However, in the process of converting a neighboring site, a WT site may `mutate', i.e., convert the site into M instead of WT with probability $\mu$. We do not consider back mutations, i.e., conversion of M sites into WT sites. The system is initialised with a linear front of WT at the bottom of the system and periodic boundary conditions are applied to the boundaries on the sides.

In this model and with this initial condition, a population of WT and M expands upwards on a lattice with regions of M forming (Fig.\ \ref{Fig1_Supplement}b). Due to the absence of back mutations, a front being composed of only M is the one and only absorbing state. How quickly this state is reached depends on the parameters of the system (Fig.\ \ref{Fig1_Supplement}b).

\section{Simulations}

\subsection{Random distribution of patches}

Throughout this work, we consider circular patch area ratio $\phi$ as control parameter. This parameter relates to the number density $n$ in an infinitely large system by \cite{xia1988,torquato2002}
\begin{equation*}
    \phi = 1-e^{-n \pi R^2}\, .
\end{equation*}

We chose the number of patches to be placed as the largest integer smaller or equal to number density times domain size. Patches were placed uniformly within the domain while taking care of periodic boundary conditions.

\subsection{Patches arranged in hexagonal lattice}

When arranging patches in a hexagonal lattice, the system width needed to be adjusted slightly to ensure periodic boundary conditions of the system.

\subsection{Emergence of mutations and conditions upon which simulations ended}

Specific observables required us to slightly modify the model and define conditions upon which a given simulation ends:
\begin{itemize}
    \item \textbf{Cluster Size Distribution}, Fig.\ 2a in main text: A single M is seeded to occur on a specific site when it is occupied by WT, and $\mu=0$ throughout the simulation. This is to ensure that only a single M cluster is monitored. The simulation ends when the front consists entirely of WT again.
    \item \textbf{Probability of patch invasion}, Fig.\ 2b in main text: For a single patch and a $\mu>0$, an initially WT population frontier engulfs a patch. If M is placed within a patch, the repeat reports an invasion success. This processes is repeated 10000 times.
    \item \textbf{Escape Region Height}, Figs.\ 2c,d in main text: Here, a single M is seeded to occur at a specific invasion angle, $\alpha$, just outside the patch. The simulation ends when there are no M on the population frontier and is repeated a 1000 times.
    \item \textbf{Probability of M dominating}, phase diagrams in Figs.\ 4 and 5 in main text. For a specific M fitness, $F$, and patch area ratio, $\phi$, over $50$ repeats the mutation rate, $\mu$, is finite until a mutation occurs, at which point $\mu$ is set to $0$. Each repeat terminates after one of the following conditions is met: \textbf{(i)} When the entire population frontier consists of M (M strain has dominated system). \textbf{(ii)} When the entire population frontier consists of WT (M strain has gone extinct). \textbf{(iii)} When the height of the population frontier exceeds a limit: in the context of a procedurally-generated vertically expanding system with fixed width, this corresponds to stopping a simulation that takes too long to end. During the simulations, the number of M on the population frontier is recorded for each time step. If the simulation terminates for either of the latter two reasons, this record is checked: if the maximum number of M on the population frontier does not exceed $2R$, this implies that no mutant ever invaded a patch. In this case, this run is discarded, and the simulation is repeated for the same patch distribution: these `sub-repeats' continue until the simulation terminates with M having invaded a patch, or some limit is reached (this limit, $50$, is to ensure that an unfortunate patch distribution does not result in a system where patches result in the WT being cut off before M can reasonably appear).
\end{itemize}

\section{Cluster size distribution in a flat front model}

Constructing an analytical description of the rough front cluster size distribution is difficult. The roughness of the front in the Eden model, characterised by standard deviation of front advancement or height, is well understood in the context of the KPZ universality class \cite{KPZ,Kuhr2011}. The implementation of mutation and selection perturbs the model away from the KPZ dynamics \cite{Kuhr2011}; to understand the effect of the M clusters, Kuhr et al.\ phenomenologically extracted the dependence of cluster correlation lengths, corresponding to their vertical and horizontal extensions, on fitness and mutation rate as well as width of the domain or lattice \cite{Kuhr2011}. We are not aware of an analytical description of the probability density function for cluster sizes.

In order to better understand the cluster size distribution, we implemented a version of our model with a flat front and without patches (Fig.\ \ref{Fig2_Supplement}a). The system is updated row-by-row within the same time step, corresponding to a linear Domany-Kinzel model \cite{domany1984}. The  simplification of a flat front allows us to obtain an analytical result for the flat front cluster height distribution, which we compare to the rough-front equivalent determined from simulations (Fig.\ 2a in main text). This height distribution of so-called isotropic clusters, clusters in which both edges have equal probabilities of moving inwards and equal probabilities of moving outwards, has been derived previously \cite{domany1984,essam1989} to be Eq.\ \ref{Eq:HeightDistribution} below. It will be re-derived here as starting point to derive the distribution of the maximum widths of clusters.

\subsection{Height Distribution}

\begin{figure}[h]
    \centering\includegraphics{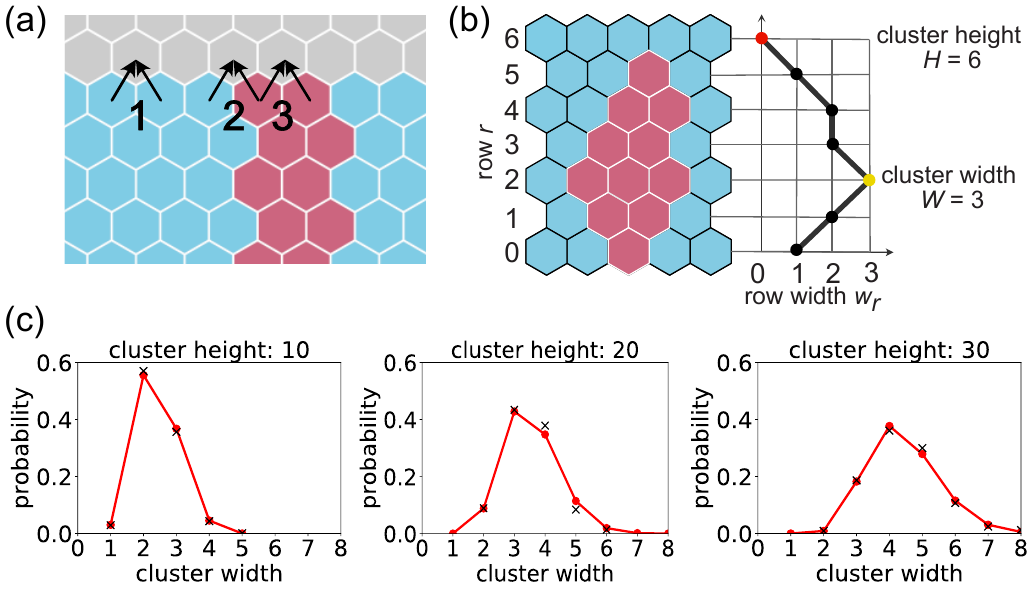}
    \caption{\textbf{(a)} Flat front model where each row is updated based on the preceding row. The fate of an unoccupied site depends only upon the two occupied neighbors below it. Three individual cases are highlighted and described in the text. \textbf{(b)} Illustration of cluster width as a random walk with initial condition $w_0=1$ and absorbing boundary condition $w=0$ (red circle). Cluster width (yellow circle) is shorthand for maximum width. \textbf{(c)} Probability of a cluster of height $H$ having width $W$ (left: $H=10$, center: $H=20$, right: $H=30$). Black crosses indicate simulations, red lines the predictions by Eq.\ \ref{Eq:ProbWidthGivenHeight}.}
    \label{Fig2_Supplement}
\end{figure}

We consider mutation rate to be so low that clusters emerge from single mutations and that no other mutations occur in the context of this cluster. A cluster of interest then starts from a flat frontier of WT with a single M site of fitness $F$. The relative probability of a site on the next row being selected to be M depends on the two neighbors it has on the current or original row, see Fig.\ \ref{Fig2_Supplement}a. One of these neighbors is to be the parent of the site in question: If both are of the same type, then the site will be the same type as well (cases 1 and 3 in Fig.\ \ref{Fig2_Supplement}a). If they are different (case 2 in Fig.\ \ref{Fig2_Supplement}a), the probability of the site being occupied by M is $\frac{F}{1+F}$, where $F$ is the relative weight that $M$ will be the parent site in analogy to the rough-front model described above.

Let us denote the width of the M cluster on row $r$, i.e., the number of M sites in that row, as $w_r$. The Markovian transition probabilities  $p_{w_r,w_{r+1}}$ (probability of the width of M on subsequent rows changing from $w_r$ to $w_{r+1}$) are:
\begin{equation}
    p_{w_r,w_{r+1}} = \begin{cases}
                            \left(\frac{F}{1+F}\right)^2 & w_r >0,\quad w_{r+1} = w_{r}+1\\
                            \frac{2F}{\left(1+F\right)^2} & w_{r} > 0,\quad w_{r+1} = w_{r}\\
                            \left(\frac{1}{1+F}\right)^2 & w_{r} > 0,\quad w_{r+1} = w_{r}-1\\
                            1 & w_{r} = w_{r+1} = 0\\
                        0 & \text{otherwise}
                            \end{cases}\, .
    \label{Eq:TransitionProbs}
\end{equation}
$p_{0,0}=1$ means that once a cluster terminates it stays terminated. The initial condition is $w_0=1$: A cluster starts from a single site originating from a mutation (Fig.\ \ref{Fig2_Supplement}b).

Note that Eq.\ \ref{Eq:TransitionProbs} is analogous to a random walk starting from $w_0=1$ with absorbing boundary condition at $0$. The  probability to step right ($w_{r+1}=w_r+1$) is  $a=\left({F}/{(1+F)}\right)^2$, to not change site ($w_{r+1}=w_r$) is $b={2F}/{\left(1+F\right)^2}$, and finally to step left ($w_{r+1}=w_r-1$) is $c=\left({1}/{(1+F)}\right)^2$.

The height $H$ of the cluster is defined as the row $r$ for which $w_r=0$ for the first time. Note that with $w_0=1$ this means there are $H$ rows that have at least one M site. A cluster of height $H$ involves $H$ random walk steps (with `stationary steps' included).

We first consider clusters that have an odd height $H=2n+1$ ($n\in\mathbb{N}^0$) and in which there are no stationary steps present (we shall define the number of stationary steps in this random walk as $S$).  The $H$th step must be a left step in order for the cluster to terminate. Given that we start with $w_0=1$ and demand $w_{2n}=1$, the remaining sequence of length $2n$ consists of left steps and right steps, but there can never be a point at which the total number of left steps outnumbers the number of right steps because that would mean the cluster terminated before reaching height $H$. This corresponds to the number of Dyck paths of length $2n$, known to be the $n^{th}$ Catalan number $C_n$ \cite{DyckPaths}:
\begin{equation*}
    C_n = \frac{1}{n+1} \binom{2n}{n}\, .
\end{equation*}

Thus, the probability of getting to state $w_H=0$ from state $w_0=1$ (with $w_{r<H}>0$) without making any stationary steps is:
\begin{equation*}
    P\left(H | S=0\right) = C_n a^{n} c^{n+1}\, .
\end{equation*}

Let us now consider the possibility of stationary steps in clusters of height $H=2n+1$. The number of left steps needs to be one larger than the number of right steps. One can only substitute pairs of left and right steps by stationary steps without changing cluster height. Thus, the number of stationary steps needs to be even. For example, a cluster of height $H=2n+1$ may consist of $n$ left steps, $n-1$ right steps, and $2$ stationary steps. The probability to obtain such a cluster is the product of the probability to have two stationary steps $p=b^2$, the probability to have a cluster of height $H=2n-1$ without stationary steps (see above), and a combinatorial factor for embedding the two stationary steps within the cluster ($\binom{2n}{2}$):
\begin{equation*}
    P\left(H|S=2\right) = \binom{2n}{2} \cdot C_{n-1} a^{n-1} c^{n} \cdot b^2 \, .
\end{equation*}

Following the same argument we obtain the probability for a cluster of height $H=2n+1$ with $2k$ stationary steps):
\begin{equation*}
    P\left(H|S=2k\right) = \binom{2n}{2k} \cdot C_{n-k} a^{n-k} c^{n-k+1} \cdot b^{2k} \, .
\end{equation*}

The total probability for the cluster to be of height $H$ is then
\begin{eqnarray*}
    P\left(H\right) & = & \sum_{k=0}^n P\left(H|S=2k\right) \\
      & = & \sum_{k=0}^n \binom{2n}{2k} \cdot C_{n-k} a^{n-k} c^{n-k+1} \cdot b^{2k} \\
      & = & a^{n} c^{n+1} C_{2n+1} \, ,
\end{eqnarray*}
where in the last step we used $b^2/ac=4$ and a computer algebra system \cite{WolframAlpha}.

Expressing the results in terms of fitness $F$ and cluster height $H$ results in
\begin{equation}
    P\left(H\right) = C_H \frac{F^{H-1}}{(1+F)^{2H}}\, .
    \label{Eq:HeightDistribution}
\end{equation}
The mean of the cluster size follows to be $\left<H\left(F\right)\right> = \frac{1+F}{1-F}$.

A similar sequence of arguments for clusters of even height $H=2n$ yields again Eq.\ \ref{Eq:HeightDistribution}. This analytical result matches the simulations for many individual clusters for fitness $F=0.9$ as shown in Fig.\ 2a of the main text. (Flat front simulations were implemented and performed independently from the Eden model described above.)

\subsection{Width Distribution}

To obtain the distribution of width $W$ for clusters of height $H$, we follow the same strategy, but now consider subsets of Dyck paths. We again first restrict our derivation to clusters of odd height, i.e., $H = 2n+1$ with $n\in\mathbb{N}^0$.

In the absence of stationary steps, a cluster of height $H$ is described by Dyck paths of length $2n$ of which there are $C_n$ ($C_n$ again being the Catalan number). Of those, the number of paths that explore a width $W$ ($1 \leq W \leq n$) from the origin is given by $T\left(n,W\right)$ \cite{oeisA080936,oeisA080934} with
\begin{eqnarray*}
    T\left(n,W\right) & = & T^{*}\left(n,W\right) - T^{*}\left(n,W-1\right),\\
    T^{*}\left(n,W\right) & = & \frac{2^{2n+1}}{W+2} \sum_{i=1}^{W+1}\left(\sin{\left(\frac{i\pi}{W+2}\right)} \left(\cos{\left(\frac{i\pi}{W+2}\right)}\right)^{n}\right)^2, n \geq 1\, .
    \label{Eq:WidthDistribution}
\end{eqnarray*}

If the Dyck path explores a width $W$, then the corresponding cluster has a maximum width of $W+1$. In the absence of stationary steps ($S=0$) the probability to obtain a cluster of width $W$ and height $H$ is thus
\begin{equation*}
    P(W, H | S=0) = \frac{T(n,W-1)}{C_n} P(H | S=0)\, ,
\end{equation*}
with $P(H | S=0)$ derived above. To obtain $P(W,H)$ we need to consider stationary steps and sum over all possible numbers of stationary steps. It is easy to see that
\begin{equation*}
    P(W, H | S=2) = \frac{T(n-1,W-1)}{C_{n-1}} P(H | S=2)\, .
\end{equation*}
Generally, the number of stationary steps needs to be even, $S=2k$, as above, where the maximum value of $k$ is n, which corresponds to a cluster of only stationary steps terminated by left step. Generalising the case of $S=2$ and summation leads to
\begin{eqnarray*}
    P(W, H) & = & \sum_{k=0}^{n-(W-1)}\frac{T(n-k,W-1)}{C_{n-k}} P(H | 2k) \\
    & = & \sum_{k=0}^{n-(W-1)}\frac{T(n-k,W-1)}{C_{n-k}}\cdot \binom{2n}{2k} C_{n-k} a^{n-k} c^{n-k+1} b^{2k}\, .
\end{eqnarray*}

Using the definitions of $a$, $b$, and $c$, substituting the index of summation to $m$ with $k=n-m$, and reversing the sum, we obtain:
\begin{eqnarray*}
    P(W, H) & = & \sum_{m=W-1}^{n} T(m,W-1) \binom{2n}{2m} a^{m} b^{2n-2m} c^{1+m}\\
            & = & b^{2n}c \sum_{m=W-1}^{n} T(m,W-1)\binom{2n}{2m} \frac{1}{2^{2m}}\\
            & = & \frac{\left(2F\right)^{H-1}}{\left(1+F\right)^{2H}}\sum_{m=W-1}^{n} T(m,W-1)\binom{H-1}{2m} \frac{1}{2^{2m}}\, .
    \label{Eq:ProbWidthHeight}
\end{eqnarray*}

Combining Eqs.\ \ref{Eq:HeightDistribution} and \ref{Eq:ProbWidthHeight}, we obtain the distribution for the width of clusters for given height:
\begin{eqnarray*}
    P\left(W \middle| H\right)  & = & \frac{P\left(W\cap H\right)}{P\left(H\right)}\\
    & = & \frac{\frac{\left(2F\right)^{H-1}}{\left(1+F\right)^{2H}}\sum_{m=W-1}^{\frac{H-1}{2}} T\left(m,W-1\right)\binom{H-1}{2m} \frac{1}{2^{2m}}}{\frac{F^{H-1}}{(1+F)^{2H}} C_H}\\
    & = & \frac{2^{H-1}}{C_H}  \sum_{m=W-1}^{\frac{H-1}{2}} T\left(m,W-1\right)\binom{H-1}{2m} \frac{1}{2^{2m}}\, .
    \label{Eq:ProbWidthGivenHeight}
\end{eqnarray*}
A similar argument for clusters with even height $H$ leads to Eq.\ \ref{Eq:ProbWidthGivenHeight} as well.

We compare this analytical expression with simulation results in Fig.\ \ref{Fig2_Supplement}b for $F=0.9$.

\section{Height of escape region as function of patch invasion angle}
\label{Sec:IsolatedPatch}

We compute the height of the escape region by equating the time it takes M to propagate through and beyond a patch with the time it takes WT to propagate around and beyond the same patch as sketched in Fig.\ \ref{Eq:EscapeRegion}.

\subsection{Invasion below the patch's horizontal symmetry axis}

\begin{figure}[H]
    \centering\includegraphics{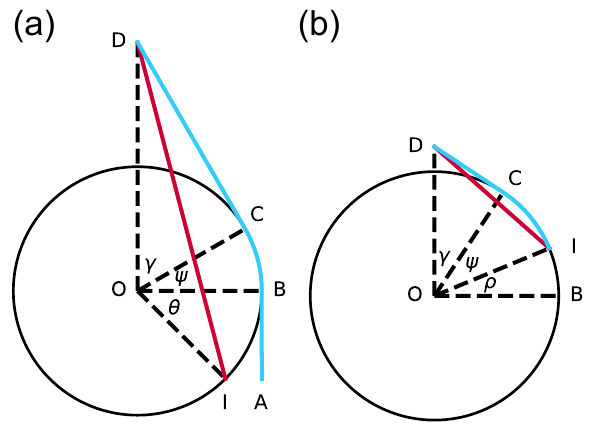}
    \caption{Paths of M through patches and WT around patches to the tip of the escape region. Patch invasion occurs at point $I$, the escape region extends up to point $D$, with escape region height $|\oline{OD}|-R$. \textbf{(a)} Invasion occurring below the patch's horizontal symmetry axis, \textbf{(b)} invasion occurring above the patch's horizontal symmetry axis.}
    \label{Eq:EscapeRegion}
\end{figure}

Let us first consider invasion of a patch (radius R, centered at origin) below the patch's horizontal symmetry axis (Fig.\ \ref{Eq:EscapeRegion}a). The location of the invasion point $I$ can be parameterized by an angle $\theta$. To determine the maximum height of the escape region, we equate the time taken by $M$ to propagate from $I$ to the tip of the escape region, $D$, with the time WT requires to reach $D$. A single patch is encompassed by a linear front below the horizontal symmetry axis and the shortest path from the initial front to the patch, $\oline{AB}$, is perpendicular to the front \cite{Moebius2015}. The path then follows the perimeter of the patch, $\wideparen{BC}$, before following the tangent from $C$ to $D$.

Note that $D$, the point M and WT reach after the same amount of time, is located on the vertical axis of symmetry, because the WT path shown in blue together with a symmetric partner on the other side cut off M at point $D$. We assume for now that $D$ is located above the patch, i.e., that an escape region exists.

The length of the path taken by M can be obtained immediately:
\begin{eqnarray*}
    \abs{\oline{ID}} = \sqrt{\left(R \cos{\theta}\right)^2 +\left(R\sin{\theta}+\abs{\oline{OD}}\right)^2 }\, .
\end{eqnarray*}

The path taken by WT is subdivided into three parts: The straight line $\oline{AB}$, the circular arc $\wideparen{BC}$, and the straight line $\oline{CD}$. Point $C$ is thereby determined by the fact that $\oline{CD}$ is tangent to the circle defining the patch. From top to bottom we obtain:
\begin{itemize}
    \item $\abs{\oline{CD}} = \sqrt{\abs{\oline{OD}}^2-R^2}$ by virtue of the right-handed triangle $OCD$,
    \item $\abs{\wideparen{BC}} = R \arcsin{\frac{R}{\abs{\oline{CD}}}}$ because $\gamma = \arccos{\frac{R}{\abs{\oline{CD}}}}$ and thus $\psi = \arcsin{\frac{R}{\abs{\oline{CD}}}}$,
    \item $\abs{\oline{AB}} = R\sin{\abs{\theta}}$.
\end{itemize}

Without loss of generality we can set the front speed of WT to 1; the front speed of M then equals fitness $F$. The time for M to reach $D$ is
\begin{equation*}
    T_{M} = \frac{1}{F} \sqrt{R^2 + 2R\abs{\oline{OD}}\sin{\theta} +\abs{\oline{OD}}^2}\, ,
\end{equation*}
the time for WT to reach $D$ is
\begin{equation*}
    T_{WT} = \sqrt{\left(R \cos{\theta}\right)^2 +\left(R\sin{\theta}+\abs{\oline{OD}}\right)^2}\, .
\end{equation*}

We can solve for the rescaled escape region height $\Lambda$ defined as $\abs{\oline{OD}}/R-1$ by equating $T_M$ with $T_{WT}$:
\begin{equation*}
    \frac{1}{F} \sqrt{1 + 2\left(1+\Lambda\right)\sin{\theta} +\left(1+\Lambda\right)^2} = \sin{\theta} + \arcsin{\frac{1}{1+\Lambda}} + \sqrt{\left(1+\Lambda\right)^2-1}\, .
\end{equation*}

In the main text we use the invasion angle $\alpha=\frac{\pi}{2} - \theta$ instead of $\theta$ resulting in
\begin{equation}
    \frac{1}{F} \sqrt{1 + 2\left(1+\Lambda\right)\cos{\alpha} +\left(1+\Lambda\right)^2} = \cos{\alpha} + \arcsin{\frac{1}{1+\Lambda}} + \sqrt{\left(1+\Lambda\right)^2-1}\, .
    \label{Eq:CapHeightDistribution_Lower}
\end{equation}

\subsection{Invasion above the patch's horizontal symmetry axis}

A relation analogous to Eq.\ \ref{Eq:CapHeightDistribution_Lower} can be derived for the case of invasion occurring above the patch's horizontal symmetry axis with M propagating from $I$ to $D$ in a straight line and WT propagating from $I$ to $D$ first along the perimeter of the circle to $C$ and then following the tangent to $D$ (Fig.\ \ref{Eq:EscapeRegion}b). We obtain:
\begin{eqnarray*}
    T_{M} & = & \frac{1}{F} \sqrt{\left(R \cos{\rho}\right)^2 +\left(R\sin{\rho}+\abs{\oline{OD}}\right)^2 }\, ,\\
    T_{WT} & = & R\left(\arcsin{\frac{R}{\abs{\oline{OD}}}} - \rho \right) + \sqrt{\abs{\oline{OD}}^2-R^2}\, .
\end{eqnarray*}
Equating both times, using rescaled escape height $\Lambda=\abs{\oline{OD}}/R-1$ and invasion angle $\alpha=\rho + \frac{\pi}{2}$, yields:
\begin{equation}
    \frac{1}{F} \sqrt{1 - 2\left(1+\Lambda\right)\cos{\alpha} +\left(1+\Lambda\right)^2} = \arcsin{\frac{1}{\left(1+\Lambda\right)}} - \alpha + \frac{\pi}{2}  + \sqrt{\left(1+\Lambda\right)^2-1}\, .
    \label{Eq:CapHeightDistribution_Upper}
\end{equation}

\subsection{Vanishing escape region height}

For some values of $F$, $\theta$, and $\rho$ there may no real positive solution to Eq.\ \ref{Eq:CapHeightDistribution_Lower} or Eq.\ \ref{Eq:CapHeightDistribution_Upper}: these correspond to cases where the M does not emerge from the patch before being cut off by WT. This corresponds to a non-existent escape region and thus $\Lambda=0$.

\subsection{Maximum escape region height}

The angle of maximum escape region height can be found by implicit differentiation. We first consider Eq.\ \ref{Eq:CapHeightDistribution_Lower}:
\begin{eqnarray*}
    \frac{\partial}{\partial\alpha} \left[\frac{1}{F} \sqrt{1 + 2\left(1+\Lambda\right)\cos{\alpha} +\left(1+\Lambda\right)^2}\right] &=& \frac{\partial}{\partial\alpha} \left[\cos{\alpha} + \arcsin{\frac{1}{1+\Lambda}} + \sqrt{\left(1+\Lambda\right)^2-1}\right]\\
    \frac{2 \frac{\partial\Lambda}{\partial\alpha}\cos{\alpha} - 2\left(1+\Lambda\right)\sin{\alpha} + 2\left(1+\Lambda\right)\frac{\partial\Lambda}{\partial\alpha}}{F \sqrt{1 + 2\left(1+\Lambda\right)\cos{\alpha} +\left(1+\Lambda\right)^2}} &=& -\sin{\alpha} - \frac{1}{\sqrt{\Lambda(1+\Lambda)}}\frac{\partial\Lambda}{\partial\alpha} + \frac{1+\Lambda}{\sqrt{(1+\Lambda)^2-1}}\frac{\partial\Lambda}{\partial\alpha}\, .
\end{eqnarray*}
With $\frac{\partial\Lambda}{\partial\alpha}=0$ we obtain
\begin{equation*}
    \frac{ - 2\left(1+\Lambda\right)\sin{\alpha}}{F \sqrt{1 + 2\left(1+\Lambda\right)\cos{\alpha} +\left(1+\Lambda\right)^2}} = -\sin{\alpha}\, ,
\end{equation*}
which is solved for $\alpha=0$, yielding
\begin{equation}
    \frac{1}{F} \left(2 + \Lambda\right) = 1 + \arcsin{\frac{1}{1+\Lambda}} + \sqrt{\left(1+\Lambda\right)^2-1}\, ,
    \label{Eq:CapHeightDistribution_Bottom}
\end{equation}
which can be solved for $\Lambda$ numerically.

No such extremum can be found for Eq.\ \ref{Eq:CapHeightDistribution_Upper}. Heuristically, we identify the extremum at $\alpha=0$ as local maximum of rescaled escape region height $\Lambda$ (see Fig.\ 3d in main text).

\section{Patches organised randomly}

We expect M to quickly dominate the front, i.e., the system to be supercritical, if escape regions overlap with patches so that large clusters of these macrostructures form spanning the system. The transition from sub-criticality to super-criticality will occur at a critical patch area ratio $\phi_c^*$, which itself depends on fitness $F$.

\begin{figure}[H]
    \centering    \includegraphics{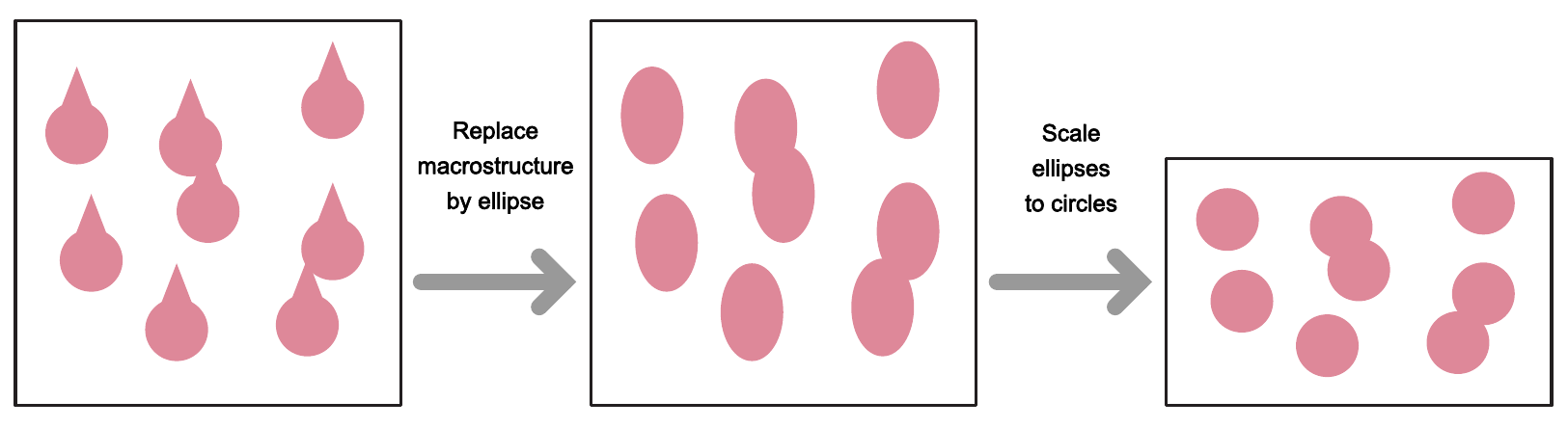}
    \caption{Description of the approach and approximation used to determine whether, for given fitness $F$ and patch area ratio $\phi$, the system displays sub-criticality (M extinction) or super-criticality (M domination). \textbf{(Left)} The patches are randomly distributed. With each patch invaded from below, fitness-dependent escape regions form. \textbf{(Center)} The macrostructure consisting of patch and escape region is approximated by an ellipse. \textbf{(Right)} Rescaling the height of the system allows one to recover the case of overlapping disks for which the percolation threshold is known. The system is sub-critical below and super-critical above the percolation threshold.}
    \label{Fig4_Supplement}
\end{figure}

Given the nature of the system with a front mainly propagating from bottom to top, we expect most invasions to occur at the bottom of the patch (invasion angle $\alpha=0$). In this case, macrostructures consist of the patch and an symmetric escape region whose height depends on fitness $F$ (Fig.\ \ref{Fig4_Supplement}, left). Without loss of generality, we can choose $R=1$, in which case escape region height is given by $\Lambda$ (Eq.\ \ref{Eq:CapHeightDistribution_Bottom}). The macrostructure thus has width 2 and height $2+\Lambda$. We approximate these macrostructures by ellipses of semi-minor axis 1 and semi-major axis $(2+\Lambda)/2$ (Fig.\ \ref{Fig4_Supplement}, centre). Note that the ellipses are aligned. Percolation is independent of rescaling the system, for example when rescaling the height of the system such that ellipses become circles (Fig.\ \ref{Fig4_Supplement}, right).

Let us denote the critical patch area ratio as $\phi_c^*$. In the thermodynamics limit, this is related to the critical number density of patches by $n_c^*$ by
\begin{equation*}
    \phi_c^* = 1-e^{-\pi n_c^*}\, ,
\end{equation*}
see, e.g., Refs. \cite{xia1988,torquato2002}.
Because the number density is independent of shape, $n_c^*$ is also the critical number density of ellipses that approximate the macrostructures. Rescaling system height by $2/(2+\Lambda)$ turns ellipses into circles. In this process, number density increases by a factor $(2+\Lambda)/2$. The critical number density $n_c^\circ$ of the discs representing the macrostructures after rescaling is therefore
\begin{equation*}
    n_c^\circ = \frac{2+\Lambda}{2} n_c^*\, ,
\end{equation*}
which in turn is related to the corresponding critical area ratio by
\begin{equation*}
    \phi_c^\circ = 1-e^{-\pi n_c^\circ}\, .
\end{equation*}

Taken together, we can express critical area ratio of patches $\phi_c^*$ assisting percolation of the slow-spreading mutants by the critical area ratio of discs in two dimensions, $\phi_c^\circ\approx 0.68$ \cite{xia1988}:
\begin{equation}
    \phi_c^*(F) = 1-\left(1-\phi_c^\circ\right)^{2/(2+\Lambda(F))} ,
    \label{Eq:Percolation}
\end{equation}
where we explicitly denote that $\phi_c^*$ depends on fitness $F$ because the (rescaled) escape region height $\Lambda$ depends on $F$. Note that Eq.\ \ref{Eq:Percolation} reduces to $\phi_c^*(F)=\phi_c^\circ$ for $\Lambda(F)=0$ as expected. When $F$ is sufficiently small such that no escape region is expected to arise, the transition line becomes vertical as seen in Fig.\ 4b of the main text.

Eq.\ \ref{Eq:Percolation} represents the predicted transition in Fig.\ 4b of the main text which matches the simulated transition very well. This may be surprising given the approximations made: (i) invasions occurring at the bottom of patches, (ii) escape regions growing deterministically, (iii) vertical alignment, (iv) approximation by ellipses. Potentially, some of approximations may cancel each other or are negligible when averaging over many patches.

\section{Patches organised in a hexagonal lattice}

When exploring the spread of M in an environment with regularly distributed patches, we consider the hexagonal lattice not only because it has the maximum patch area ratio $\phi$ of all the Bravais lattices \cite{chang2010simple}, but it also has the most degrees of symmetry. In the hexagonal lattice, the separation between the surfaces of adjacent patches, $S$, is related to the patch area ratio by $\phi={2\pi R^2}/{(\sqrt{3}\left(2R+S\right)^2)}$.

\begin{figure}[H]
    \centering    \includegraphics{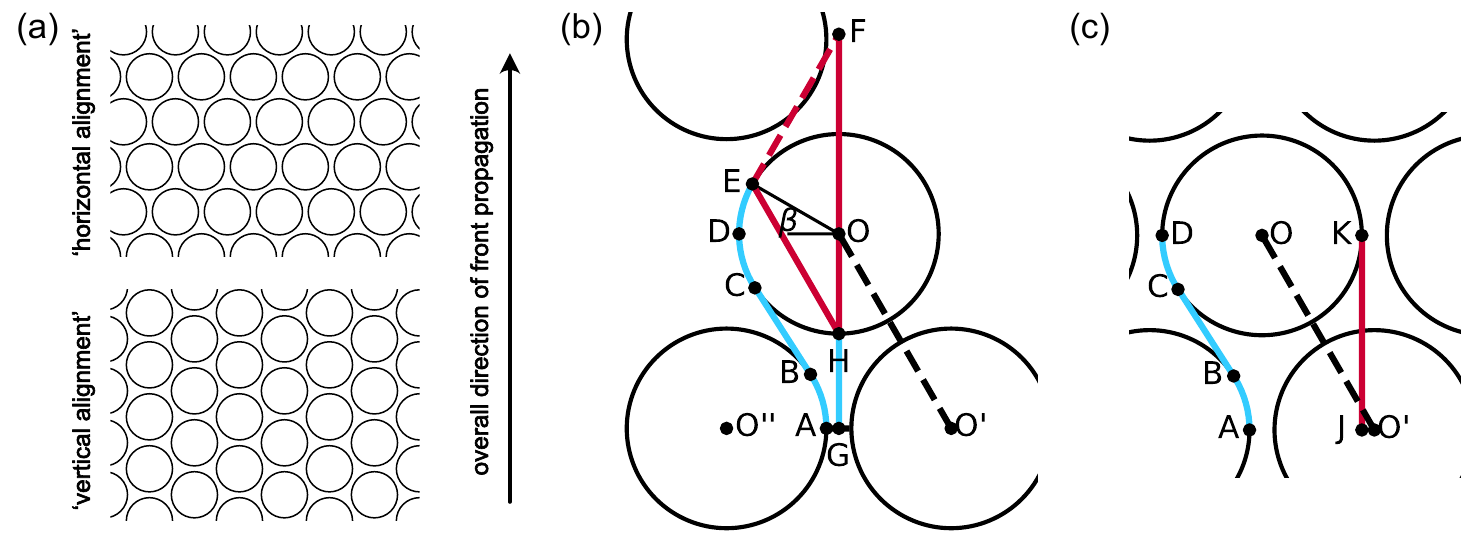}
    \caption{\textbf{(a)} Sketch of hexagonal lattice of patches aligned horizontally (upper) and vertically (lower). \textbf{(b)} Paths of WT (cyan) and M (red) taken in the short-term estimate of whether M will dominate the front, for details see Section \ref{Sec:ShortTermSuccessM}. \textbf{(c)} Paths of WT (cyan) and M (red) taken in the long-term estimate of whether M will dominate the front, for details see Section \ref{Sec:LongTermSuccessM}.}
    \label{Fig5_Supplement}
\end{figure}

Postulating that a critical area ratio $\phi_c^{\hexagon}$ exists, one can easily obtain two bounds. First, for $S=0$ neighboring patches are in contact and there is no path for WT to propagate vertically and therefore:
\begin{equation*}
    \phi_c^{\hexagon} \le \phi\left(S=0\right)=\frac{\pi}{2\sqrt{3}}\approx 0.91\, .
\end{equation*}
Secondly, when WT can propagate vertically unhindered by obstacles, isolated M clusters will become enclosed by WT for any $F<1$. The presence of these vertical paths depends upon the orientation of the lattice relative to the population front. For the case where nearest-neighbor patches are horizontally adjacent (the `horizontal alignment', as seen in Fig.\ \ref{Fig5_Supplement}a, upper), vertical channels occur for $S\ge 2R$. For the case where nearest neighbor patches are vertically adjacent (the `vertical alignment', as seen in Fig.\ \ref{Fig5_Supplement}a, lower) vertical channels are present for $S \ge \left(\frac{4}{\sqrt{3}} - 2\right)R$. Thus:
\begin{eqnarray*}
    \frac{\pi}{8\sqrt{3}} \le \phi_c^{\hexagon}\quad \textrm{for horizontal alignment}\, ,
    \frac{\pi \sqrt{3}}{8} \le \phi_c^{\hexagon}\quad \textrm{for vertical alignment}\, .
\end{eqnarray*}
In the following we only consider the horizontal alignment of the lattice of patches.

We describe two approaches for more accurately estimating the critical patch area ratio $\phi_c^{\hexagon}$ above which M dominates quickly: First, a `short-term' consideration asking if invasion of a single patch leads to the invasion of an immediate downstream patch. Second, a `long-term' consideration in which we compare the speed at which WT propagates around patches with the speed by which M propagates vertically. Note that $\phi_c^{\hexagon}$ is a function of fitness $F$; conversely, we can define a critical fitness $\phi_c^{\hexagon}$ depending on patch area ratio $\phi$ or surface separation $S$.

\subsection{Short-term success of M}
\label{Sec:ShortTermSuccessM}

We here ask whether invasion of a patch by M (from the base of the patch) will lead to a patch invasion in the next row of patches. To that end, we consider whether the escape region envelope overlaps with the patch in the next row as illustrated in Fig.\ \ref{Fig5_Supplement}b. We thereby approximate the envelope by connecting the points $E$ (where WT and M arrive at the same time at the patch's boundary) and $F$ (the point determining escape region height considered above). Note that this is based on the observation that the escape region has the shape of a triangle, rather than a rigorous derivation of escape region shape.

As starting condition, we consider the WT front to be approximately flat along the line $\oline{AG}$ between the patches. Previous work \cite{Moebius2015,Moebius2021} predicts that the front would be perturbed due to patches below; we neglect this effect as well as stochasticity-induced roughness.

Let us first derive the location of the point $E$ parameterized by angle $\beta=\angle DOE$. The paths for WT to reach point E from point A can be decomposed into four segments with (arc) lengths as follows:
\begin{eqnarray*}
    \abs{\oline{BC}} & = & 2\sqrt{\left(R+\frac{S}{2}\right)^2-R^2} \quad \textrm{(length of inner tangent of two neighboring patches)}, \\
    \abs{\wideparen{AB}} & = & R\left(\frac{\pi}{3}-\arccos{\frac{R}{R+\frac{S}{2}}}\right) \quad \textrm{(length of circular arc, difference of $\angle O'O''O$ and $\angle BO''O$)}, \\
    \abs{\wideparen{CD}} & = & \abs{\wideparen{AB}} \quad \textrm{(because of symmetry)}, \\
    \abs{\wideparen{DE}} & = & \beta R \quad \textrm{(length of circular arc)}.
\end{eqnarray*}

M reaches point E via a different path, first propagating as WT to point H, mutating, and following the secant to point $E$. The corresponding paths lengths are:
\begin{equation*}
    \abs{\oline{GH}} = \frac{\sqrt{3}}{2}\left(2R+S\right) - R \quad \textrm{and} \quad \abs{\oline{HE}} = R\sqrt{2+2\sin{\beta}}.
\end{equation*}

Equating the time to reach E, taking into account that M is propagating with relative speed $F$, we yield:
\begin{eqnarray}
    \abs{\oline{BC}} + \abs{\wideparen{AB}} + \abs{\wideparen{CD}} + \abs{\wideparen{DE}} & = & \abs{\oline{GH}} + \frac{1}{F}\abs{\oline{HE}}\, , \\
    \sqrt{\left(R+\frac{S}{2}\right)^2-R^2} + 2\left(\frac{\pi}{3}-\arccos{\frac{R}{R+\frac{S}{2}}}\right) + R\beta & = & \frac{\sqrt{3}}{2}\left(2R+S\right) - R + \frac{1}{F} R\sqrt{2+2\sin{\beta}}\, . \nonumber
    \label{Eq:Lattice_TimeEmergence}
\end{eqnarray}

To find F, located at the tip of the escape region on the axis of symmetry, we perform a similar procedure as in the case of an isolated patch (Section \ref{Sec:IsolatedPatch}), but this time taking into account the effect of neighboring patches, i.e., the less direct path from A to D:
\begin{eqnarray}
    \abs{\wideparen{AB}}+\abs{\oline{BC}}+\abs{\wideparen{CD}}+\abs{\wideparen{DE}}+\frac{1}{F}\abs{\oline{EF}} & = & \abs{\oline{GH}} + \frac{1}{F}\abs{\oline{HF}}\, , \\
    \sqrt{\left(R+\frac{S}{2}\right)^2-R^2} + 2\left(\frac{\pi}{3}-\arccos{\frac{R}{R+\frac{S}{2}}}\right) + R\beta + \frac{1}{F} \abs{\oline{EF}} & = & \frac{\sqrt{3}}{2}\left(2R+S\right) - R + \frac{1}{F}\left(R+\abs{\oline{OF}}\right)\, . \nonumber
    \label{Eq:Lattice_TimeTip}
\end{eqnarray}

For given fitness $F$ and separation $S$, solving Eq.\ \ref{Eq:Lattice_TimeEmergence} numerically for $\beta$ and subsequently Eq.\ \ref{Eq:Lattice_TimeTip} for $\abs{\oline{OF}}$ determines both the location of E and F. We then determine numerically whether $\oline{EF}$ intersects with the neighboring patch as is the case in Fig.\ \ref{Fig5_Supplement}b.

In summary, making a number of assumptions outlined, we can predict whether invasion of a patch leads to invasion of a neighboring patch downstream for given fitness $F$ and surface separation $S$ (or equivalently patch area ratio $\phi$), leading to critical area ratio $\phi_c^{\hexagon}(F)$ or critical fitness $F_c^{\hexagon}(\phi)$.

\subsection{Long-term success of M}
\label{Sec:LongTermSuccessM}

The analysis above does not take into account the long-term time evolution of the system. If M spreads from one patch to another downstream, M does not necessarily need to invade further patches because in the next row the invasion occurs at a different invasion angle (from the bottom right instead of the bottom in Fig.\ \ref{Fig5_Supplement}b). This motivates the need to consider a more robust evaluation with focus on the long-term time evolution of the system, sketched in  Fig.\ \ref{Fig5_Supplement}c, which asks whether WT following the winding paths around patches or M propagating through patches with slower relative speed $F$ are overall faster.

The path length of WT from one row to the next around patches is circular segments $\wideparen{AB}$ and $\wideparen{CD}$ as well as the segment $\oline{BC}$, which forms part of the inner tangent. The length of these paths has been considered above and we obtain:
\begin{equation*}
    \abs{\wideparen{AB}}+\abs{\oline{BC}}+\abs{\wideparen{CD}} = \sqrt{\left(R+\frac{S}{2}\right)^2-R^2} + 2R\left(\frac{\pi}{3} - \arccos\frac{2R}{2R+S}\right)\, .
\end{equation*}

The length of the direct path from one row to the next is given by
\begin{equation*}
    \abs{\oline{JK}} = \frac{\sqrt{3}}{2} (2R+S)\, .
\end{equation*}

WT can outpace the M despite the longer path length if
\begin{equation*}
    \abs{\wideparen{AB}}+\abs{\oline{BC}}+\abs{\wideparen{CD}} < \frac{1}{F} \abs{\oline{JK}}\, .
\end{equation*}
Thus, the critical fitness $F_c^{\hexagon LT}$, depending on surface separation $S$ using the long-term consideration, is
\begin{equation*}
    F_c^{\hexagon LT}(S) = \frac{\sqrt{3}}{2} \frac{2R+S}{\sqrt{S^2 +4RS} + 2R\left(\frac{\pi}{3} - \arccos\frac{2R}{2R+S}\right)}\, .
\end{equation*}

\section{Captions for Supplementary Videos}

\textbf{Video 1 (relating to Fig 1c):}
Time evolution of the modified Eden model, with initially flat front of WT in a domain of $1000\times 1000$ lattice points. When WT invades an empty site, it has a mutation probability $\mu = 0.001$ of mutating into M of fitness $F=0.95$. Only M are able to invade the circular patches of radius $R=50$. Patches are placed such that about half of the system is covered by patches $\phi=0.5$. The system is expected to result in rapid M domination of the front if $F\geq1$ or $\phi > \approx 0.68$, but here we see rapid M domination in spite of neither of these conditions being satisfied.

\textbf{Video 2 (relating to Fig 2ci and 3ai):}
Time evolution of the modified Eden model, with initially flat front of WT in a domain of $400\times 400$ lattice points. When WT invades an empty site, it has a mutation probability $\mu = 0.0005$ of mutating into M of fitness $F=0.9$. Only M are able to invade the circular patch of radius $R=50$. Here, a cluster of M invades the patch from the bottom, resulting in an `escape region' forming vertically above the patch.

\textbf{Video 3 (relating to Fig 2cii and 3aii):}
Time evolution of the modified Eden model, with initially flat front of WT in a domain of $400\times 400$ lattice points. When WT invades an empty site, it has a mutation probability $\mu = 0.0005$ of mutating into M of fitness $F=0.9$. Only M are able to invade the circular patch of radius $R=50$. Here, a cluster of M invades the patch a significant distance away from the bottom of the patch; this results in an `escape region' forming vertically above the patch, which is significantly smaller than the one in Fig.\ 2ci (Video S2).

\textbf{Video 4 (relating to Fig 4ai):}
Time evolution of the modified Eden model, with initially flat front of WT in a domain of $2000\times 2000$ lattice points. When WT invades an empty site, it has a mutation probability $\mu = 0.001$ of mutating into M of fitness $F=0.9$. Only M are able to invade the circular patches of radius $R=40$. Patches are randomly placed such that patch area ratio $\phi\approx 0.4$. WT persists on the population frontier for the entirety of the simulation, indicating that the system is sub-critical for this parameter pair $(F,\phi)$.

\textbf{Video 5 (relating to Fig 4aii):}
Time evolution of the modified Eden model, with initially flat front of WT in a domain of $2000\times 2000$ lattice points. When WT invades an empty site, it has a mutation probability $\mu = 0.001$ of mutating into M of fitness $F=0.98$. Only M are able to invade the circular patches of radius $R=40$. Patches are randomly placed such that patch area ratio $\phi\approx 0.4$. M rapidly dominates the population frontier, indicating that the system is super-critical for this parameter pair $(F,\phi)$.

\textbf{Video 6 (relating to Fig 4aiii):}
Time evolution of the modified Eden model, with initially flat front of WT in a domain of $2000\times 2000$ lattice points. When WT invades an empty site, it has a mutation probability $\mu = 0.001$ of mutating into M of fitness $F=0.9$. Only M are able to invade the circular patches of radius $R=40$. Patches are randomly placed such that patch area ratio $\phi\approx 0.7$. M rapidly dominates the population frontier, indicating that the system is super-critical for this parameter pair $(F,\phi)$. Note that this is because patches are in the continuum percolation regime, creating a spanning cluster of patches that head off WT.

\textbf{Video 7 (relating to Fig 5ai):}
Time evolution of the modified Eden model, with initially flat front of WT in a domain of $2000\times 2000$ lattice points. When WT invades an empty site, it has a mutation probability $\mu = 0.001$ of mutating into M of fitness $F=0.9$. Only M are able to invade the circular patches of radius $R=40$. Patches are placed on a hexagonal lattice. The surface separation of the nearest-neighbor patches are chosen such that patch area ratio $\phi=0.4$. WT persists on the population frontier, with M clusters being finite, indicating that the system is sub-critical for this parameter pair $(F,\phi)$.

\textbf{Video 8 (relating to Fig 5aii):}
Time evolution of the modified Eden model, with initially flat front of WT in a domain of $2000\times 2000$ lattice points. When WT invades an empty site, it has a mutation probability $\mu = 0.001$ of mutating into M of fitness $F=0.98$. Only M are able to invade the circular patches of radius $R=40$. Patches are placed on a hexagonal lattice. The surface separation of the nearest-neighbor patches are chosen such that patch area ratio $\phi=0.4$. We see that M rapidly dominates the population frontier, indicating that the system is super-critical for this parameter pair $(F,\phi)$

\textbf{Video 9 (relating to Fig 5aiii):}
Time evolution of the modified Eden model, with initially flat front of WT in a domain of $2000\times 2000$ lattice points. When WT invades an empty site, it has a mutation probability $\mu = 0.001$ of mutating into M of fitness $F=0.9$. Only M are able to invade the circular patches of radius $R=40$. Patches are placed on a hexagonal lattice. The surface separation of the nearest-neighbor patches are chosen such that patch area ratio $\phi=0.8$. We see that M rapidly dominates the population frontier, indicating that the system is super-critical for this parameter pair $(F,\phi)$

\printbibliography